\documentclass[hidelinks]{article}

\usepackage{arxiv}

\usepackage[utf8]{inputenc} % allow utf-8 input
\usepackage[T1]{fontenc}    % use 8-bit T1 fonts

\usepackage{amsmath, amssymb} % Math symbols

\usepackage{hyperref}       % hyperlinks
\usepackage{url}            % simple URL typesetting
\usepackage{amsfonts}       % blackboard math symbols
\usepackage{nicefrac}       % compact symbols for 1/2, etc.
\usepackage{microtype}      % microtypography
\usepackage{graphicx}
\usepackage{doi}
\usepackage{natbib} % for citation
\usepackage{pdflscape} % for landscape view
\usepackage{acro} % for abbreviations
\usepackage{bookmark} % Adds PDF bookmarks
\usepackage{nameref} % Allows to reference not only the number but also the name of a section

\usepackage{tabularx}
\usepackage{booktabs}
\usepackage{caption}
\usepackage{tabu}
\usepackage{rotating} % landscape tables
\usepackage{makecell} % multiline cells
\usepackage{multirow} % multiline rows
\usepackage{float} % allows putting an [H] in \begin{figure} to specify the exact location of the figure
\usepackage{array}
\usepackage{arydshln}
\newcolumntype{P}[1]{>{\centering\arraybackslash}p{#1}}
\newcolumntype{R}[1]{>{\raggedleft\arraybackslash}p{#1}}

\usepackage{chngcntr}
\counterwithin{figure}{section}
\counterwithin{table}{section}
\counterwithin{equation}{section}

\usepackage[dvipsnames]{xcolor} % For colouring text

\usepackage{algorithm}
\usepackage{algpseudocode}

\usepackage{bm} % For bold letter in math mode

\usepackage{amsthm}

\title{The State of Generative AI in Software Development: Insights from Literature and a Developer Survey}

\author{
    \href{https://orcid.org/0009-0002-1502-3670}
    {\includegraphics[scale=0.06]{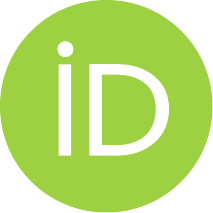}
    Vincent Gurgul} \\
	Chair of Information Systems\\
	Humboldt-Universität zu Berlin\\
	Unter den Linden 6, 10117 Berlin\\
	\And
	\href{https://orcid.org/0000-0003-4304-2123}
    {\includegraphics[scale=0.06]{orcid.pdf}
	Robin Gubela} \\
	Chair of Information Systems\\
	Hochschule für Technik und Wirtschaft Berlin\\
	Treskowallee 8, 10318 Berlin\\
	\And
	\href{https://orcid.org/0000-0001-7685-262X}
    {\includegraphics[scale=0.06]{orcid.pdf}
    Stefan Lessmann} \\
	Chair of Information Systems\\
	Humboldt-Universität zu Berlin\\
	Unter den Linden 6, 10117 Berlin\\
	Bucharest University of Economic Studies\\
    6 Piata Romana, 010374, Romania\\
}

\renewcommand{\headeright}{}
\renewcommand{\shorttitle}{The State of Generative AI in Software Development: Insights from Literature and a Developer Survey}

\hypersetup{
pdftitle={The State of Generative AI in Software Development: Insights from Literature and a Developer Survey},
pdfsubject={generative ai, software development},
pdfauthor={Gurgul, V., Gubela, R., Lessmann, S.},
pdfkeywords={generative ai, software engineering, large language models, ai-assisted coding, it governance, agile development},
}

\DeclareAcronym{GenAI}{
	short = GenAI,
	long = Generative Artificial Intelligence,
}
\DeclareAcronym{LLM}{
	short = LLM,
	long = Large Language Model,
}
\DeclareAcronym{SDLC}{
	short = SDLC,
	long = Software Development Life Cycle,
}
\DeclareAcronym{IDE}{
  short = IDE,
  long = Integrated Development Environment,
}
\DeclareAcronym{TAM}{
  short = TAM,
  long = Technology Acceptance Model,
}

\begin{document}
\maketitle

\begin{abstract}

Generative Artificial Intelligence (GenAI) rapidly transforms software engineering, yet existing research remains fragmented across individual tasks in the Software Development Lifecycle. This study integrates a systematic literature review with a survey of 65 software developers. The results show that GenAI exerts its highest impact in design, implementation, testing, and documentation, where over 70 \% of developers report at least halving the time for boilerplate and documentation tasks. 79 \% of survey respondents use GenAI daily, preferring browser-based Large Language Models over alternatives integrated directly in their development environment. Governance is maturing, with two-thirds of organizations maintaining formal or informal guidelines. In contrast, early SDLC phases such as planning and requirements analysis show markedly lower reported benefits. In a nutshell, GenAI shifts value creation from routine coding toward specification quality, architectural reasoning, and oversight, while risks such as uncritical adoption, skill erosion, and technical debt require robust governance and human-in-the-loop mechanisms.

\end{abstract}


% -----------------------------------------------------------------------------
% MAIN BODY OF TEXT
% -----------------------------------------------------------------------------

\section{Introduction}
\label{sec:introduction}

The rapid adoption of \ac{GenAI}, particularly \acp{LLM}, is substantially transforming the \ac{SDLC}. \cite{chui_economic_2023} identify software engineering as one of the domains with the highest economic impact potential from GenAI.
AI-assisted tools such as GitHub Copilot are increasingly embedded in professional software engineering environments, supporting activities that span the entire lifecycle—from requirements analysis to architectural design, code generation, testing, debugging, documentation, and maintenance.

\ac{GenAI} refers to a class of machine learning systems capable of producing novel content (such as text, images, code, or sound) by learning patterns from large datasets and generating contextually coherent outputs \citep{goodfellow_deep_2016}.
Prominent examples include LLMs such as ChatGPT, which process and generate natural language.
While these systems do not replace human judgment or accountability, they might augment developer capabilities, potentially enabling faster iteration cycles, improved efficiency, and higher software quality.
At the same time, its adoption raises questions related to maintainability, security, intellectual property, transparency, regulatory compliance, and long-term developer skill development.

Despite growing interest, the literature remains fragmented: Studies typically focus on aspects related to isolated \ac{SDLC} phases, while comprehensive assessments across the full lifecycle with empirical data remain scarce.
Research agendas such as \cite{feuerriegel_generative_2024} call for more empirical investigations into how \ac{GenAI} transforms organizational work practices and, specifically, how such systems affect software development productivity. To address these gaps, we derive the following research questions.

\noindent\begin{enumerate}
  \item[RQ1:] How are \ac{GenAI} tools currently used by software developers?
  \item[RQ2:] How is the impact of \ac{GenAI} perceived across the phases of the \ac{SDLC}?
  \item[RQ3:] What governance mechanisms are implemented to manage the adoption of \ac{GenAI} in software development departments?
  \item[RQ4:] What are the perceived risks associated with \ac{GenAI} in software development?
\end{enumerate}

This study addresses these questions by integrating three complementary sources of evidence.
First, a structured review of academic literature synthesizes current scholarly knowledge on the potential and limitations of \ac{GenAI} across the phases of the \ac{SDLC} and in agile development.
Second, industry reports are incorporated to capture emerging practices, technological developments, and practitioner-oriented assessments that often precede academic publication.
Third, a cross-sectional survey of 65 software developers building on the \ac{TAM} \citep{davis_perceived_1989, venkatesh_theoretical_2000} provides primary empirical evidence on real-world tool usage, perceived productivity effects, governance maturity, and associated risks.
By integrating insights from academic research, industry discourse, and practitioner experience, the study enables a more comprehensive and empirically grounded assessment of how \ac{GenAI} is reshaping development practices, influencing role profiles, and altering governance requirements.

\section{Methodology for Literature Analysis}
\label{sec:methodology}

This section outlines the methodology used to identify and select relevant studies for the literature analysis presented in Section \ref{sec:literature_review}.
Our literature selection process follows established methodological guidelines for conducting systematic literature reviews in Information Systems research \citep{webster_analyzing_2002,okoli_guide_2015}.
Following the typology proposed by \cite{pare_synthesizing_2015}, who distinguish nine literature review types, ours can be characterized as a scoping review:
Its primary goal is to identify and structure the current body of research on \ac{GenAI} across the SDLC, rather than to aggregate quantitative data or to develop new theoretical propositions.
By organizing the literature along the \ac{SDLC} phases, we follow the concept-centric approach suggested by \cite{webster_analyzing_2002}, shifting the focus from what individual authors have found to what is collectively known about each stage of the development process.

The selection criteria included peer-reviewed journal articles, conference proceedings, preprints, and technical reports published by established consulting and research organizations.
Extending the evidence base beyond academic publications is in line with \cite{okoli_guide_2015}, who acknowledges non-peer-reviewed sources as part of the searchable evidence base in systematic reviews.
Including such literature is relevant for our research in the rapidly evolving research field of GenAI, where much novel knowledge is not yet reflected in peer-reviewed publications.
Furthermore, only publications released between 2021 and 2026 were considered, reflecting the period during which large-scale \ac{GenAI} systems became widely accessible.

The primary databases searched were Scopus, IEEE Xplore, the ACM Digital Library, SpringerLink, arXiv, and Google Scholar.
These sources were chosen to ensure broad coverage of software engineering, information systems, and applied AI research.
Search queries combined general and domain-specific keywords.
Core search terms included “AI-assisted coding”, “GitHub Copilot”, “Generative AI”, “large language models”, “agile framework” and “software development tools”. These were complemented by phase-specific terms corresponding to the SDLC, such as “requirements engineering”, “software design”, “code generation”, “testing”, “debugging”, “maintenance” and “project management”.
In addition, selected industry reports from consulting and technology firms such as McKinsey, Boston Consulting Group (BCG), Deloitte, and Accenture were included, as these publications often provide early insights into technological adoption patterns and practical challenges that precede academic research.

The screening and selection process followed the three-stage framework of \cite{levy_systems_2006} following prior literature screening approaches.
In the first stage (input), the initial search across all databases yielded 959,044 publications.
In the second stage (processing), duplicate records across databases, non-English manuscripts, and publications prior to 2019 were removed, and results were filtered by publication type following established literature screening procedures \citep[for example][]{frohnel_facing_2024}.
This step reduced the corpus to 1,712 candidate publications.
The remaining records were then screened in two stages: first, titles and abstracts were reviewed for topical relevance, resulting in 160 candidate publications. Second, the remaining articles underwent full-text screening.
Selection criteria included relevance to the impact of GenAI on one or more SDLC phases, methodological transparency, and substantive contribution to understanding technical, organizational, or managerial implications.
Studies focusing solely on unrelated AI applications or lacking sufficient methodological clarity were excluded.
We identified additional publications through backward search \citep{webster_analyzing_2002} by reviewing the reference lists of the remaining studies and applying the same selection criteria as before.
The screening by full-text resulted in 63 publications included in the review.
In the third stage, the final corpus of publications was organized according to global observations, individual \ac{SDLC} phases, agile methodologies, and risk analysis, as presented subsequently.

\section{Literature Review}
\label{sec:literature_review}

The rapid advancement of \ac{GenAI} is increasingly transforming software development practices across the \ac{SDLC}. Among AI-assisted development tools, conversational LLMs such as ChatGPT and assistants integrated directly in the development environment like GitHub Copilot, Amazon CodeWhisperer, JetBrains AI Assistant, and Tabnine have become widely adopted, supporting tasks such as code generation, explanation, and review \citep{sergeyuk_using_2025}.

Empirical studies and field research consistently report productivity gains across development activities, particularly in code optimization (approximately 60 \%), bug fixing (26 \%), and documentation support (12 \%) \citep{collante_impact_2025}.
Evidence further suggests that less experienced developers benefit disproportionately from AI assistance, achieving higher relative productivity improvements \citep{brynjolfsson_generative_2023,dellacqua_navigating_2023,ng_harnessing_2024}.
At the same time, studies indicate that code quality generally remains stable or improves despite reduced development time \citep{shihab_effects_2025,yadav_evaluating_2025}.
However, correctness cannot be guaranteed and long-term risks remain, making human validation and oversight essential \citep{yetistiren_evaluating_2023,atif_celloai_2025}.
Beyond individual productivity, industry reports highlight broader organizational implications, including accelerated prototyping, shorter time-to-market, and shifting accountability toward product and project leadership \citep{deniz_unleash_2023, gnanasambandam_how_2025}.
The following subsections synthesize the literature across the phases of the \ac{SDLC}, before addressing agile software development as well as the risks and roadblocks associated with \ac{GenAI} adoption.

\subsection{\ac{GenAI} in the Software Development Lifecycle}

\subsubsection{Planning.} In the planning phase, \ac{GenAI} supports the creation of foundational project artifacts such as scope definitions, timelines, milestone structures, role descriptions, communication plans, and risk assessments \citep{barcaui_who_2023, hughes_impact_2025, maggoo_next_2025}.
AI systems are also used to conduct market analyses, synthesize competitive information, and formulate product goals and visions \citep{lechner_can_2025, dellacqua_navigating_2023}.
Cost estimation and story point approximation have likewise been explored as application areas \citep{wagner_evolution_2024}.

Empirical evidence suggests considerable productivity gains during project initiation and planning, including faster setup of new projects and improved alignment between technical teams and management \citep{aramali_generative_2025, brynjolfsson_generative_2023, hughes_impact_2025}.
GenAI is typically deployed via structured prompts and increasingly integrated into project management tools such as Jira, Confluence, or Microsoft Project \citep{barcaui_who_2023,ng_harnessing_2024}.
These capabilities contribute to early identification of risks and structured project preparation.

\subsubsection{Requirements Analysis.} During requirements analysis, \ac{GenAI} assists in drafting, refining, and validating functional and technical requirements.
LLMs are used to generate and elaborate user stories and use cases based on high-level project documentation \citep{cico_ai-assisted_2023, pinto_developer_2023, wei_ai-inspired_2025} and to create preliminary architectural drafts and structured feature specifications \citep{glever_redefining_2024, gong_language_2025, nitin_using_2024}.
AI-assisted approaches reduce conceptualization time and accelerate the creation of prototypes, including visually oriented UI/UX artifacts \citep{wei_requirements_2024, wei_ai-inspired_2025, gnanasambandam_how_2025}.
Tools such as Figma AI and related plugins seem promising implementations for rapid prototyping \citep{petridis_promptinfuser_2024}.
Furthermore, AI systems can analyze requirement documents to identify inconsistencies, ambiguities, and missing elements, enabling earlier error detection and reducing downstream rework \citep{gong_language_2025, porter_requirements_2025}.
Improved consistency between business and IT stakeholders has also been reported \citep{pinto_developer_2023, brynjolfsson_generative_2023}.

\subsubsection{Design and Implementation.} The design and implementation phase represents the most extensively studied application domain for GenAI.
AI-assisted coding tools such as GitHub Copilot provide real-time code completion, refactoring suggestions, boilerplate generation, and context-aware optimization \citep{peng_impact_2023,yetistiren_evaluating_2023, solohubov_accelerating_2023, glever_redefining_2024, gerdemann_why_2024}.
They also support automatic documentation generation, including code comments and API documentation \citep{atif_celloai_2025,pinto_developer_2023, ndiaye_generative_2025}, as well as recommendations to improve code structure and maintainability \citep{gong_language_2025,zhong_bepilot_2025}.

Experimental and field studies consistently report increased development speed \citep{peng_impact_2023, cui_productivity_2024, shihab_effects_2025, struever_dev_2025,deniz_unleash_2023}.
Routine tasks are particularly affected, with automation reducing manual effort for standardized coding activities \citep{solohubov_accelerating_2023}.
Several analyses also indicate improvements in review efficiency and defect reduction, although human oversight remains essential \citep{collante_impact_2025,yetistiren_evaluating_2023, ndiaye_generative_2025}.
Implementation typically occurs via integration into \acp{IDE} such as Visual Studio Code or IntelliJ, often combined with secure APIs or isolated models to address data protection and compliance requirements \citep{ng_harnessing_2024, ebert_generative_2023}.

\subsubsection{Testing and Integration.}  In the testing and integration phase, \ac{GenAI} is used to automatically generate unit, integration, and end-to-end test cases \citep{atif_celloai_2025, glever_redefining_2024, maggoo_next_2025,zhong_bepilot_2025}.
Additional applications include the creation of mock objects and synthetic test data \citep{yetistiren_evaluating_2023,gong_language_2025}, suggestions for missing test paths and coverage gaps \citep{cico_ai-assisted_2023, nitin_using_2024}, and AI-assisted security analysis \citep{ding_generative_2024, pearce_asleep_2025, ndiaye_generative_2025}.
The literature documents reductions in manual testing effort for routine scenarios and broader, faster test coverage \citep{kathiresan_automated_2024,solohubov_accelerating_2023,zhong_bepilot_2025} as well as earlier detection of edge cases and integration issues \citep{yetistiren_evaluating_2023, kathiresan_automated_2024}.
In practice, \ac{GenAI} is often integrated into CI/CD pipelines and connected to test frameworks or embedded directly into development environments for automated code analysis and test case generation \citep{ng_harnessing_2024,pinto_developer_2023}.

\subsubsection{Operation and Maintenance.} In the operation and maintenance phase, \ac{GenAI} supports the analysis of logs, error messages, and stack traces, providing diagnostic insights and suggesting potential remediation steps based on learned patterns \citep{ng_harnessing_2024,gong_language_2025,zhong_bepilot_2025}.
It is also used to explain, summarize, and restructure poorly documented or complex legacy code, improving maintainability and knowledge transfer \citep{nitin_using_2024,atif_celloai_2025, maggoo_next_2025}.

Reported effects include faster issue detection and resolution, improved traceability, and enhanced resilience to staff turnover due to automated documentation and knowledge preservation \citep{ng_harnessing_2024,pinto_developer_2023,nitin_using_2024}.
Integration commonly occurs within incident and log management platforms such as Jira, ServiceNow, or Grafana, and in sensitive environments through locally deployed or access-controlled AI systems to mitigate compliance risks \citep{ng_harnessing_2024,ebert_generative_2023}.

\subsection{GenAI in Agile Software Development}

\ac{GenAI} influences the dynamics of agile development practices beyond productivity gains. The literature reports high perceived usefulness and satisfaction in agile teams when using AI-assisted tools \citep{geyer_case_2025} and accelerated task completion, leading to shorter iteration cycles and more frequent releases \citep{zhang_empowering_2024, ulfsnes_transforming_2024, bahi_integrating_2024}.
By automating routine coding, documentation, and testing tasks, \ac{GenAI} enables teams to focus on value-generating increments, reinforcing core agile principles such as rapid feedback and continuous delivery.
At the same time, role expectations shift: Product owners, developers, and scrum masters increasingly engage in strategic, creative, and evaluative activities, while routine tasks are partially delegated to AI systems \citep{diebold_backlogs_2025, bahi_integrating_2024, nasir_automating_2024}.
The literature recommends adapting role profiles and strengthening AI-related competencies, particularly critical evaluation skills in AI-supported environments.

\subsection{Risks and Roadblocks in the Adoption of \ac{GenAI}}

Despite substantial productivity gains, the literature identifies a range of short- and long-term risks associated with the integration of \ac{GenAI} into software development.
In the short term, concerns relate to reliability, quality control, and security. AI-generated code suggestions are frequently erroneous, with manual correction efforts averaging approximately ten minutes per identified defect \citep{yetistiren_evaluating_2023}.
A significant proportion of outputs requires post-editing before productive use \citep{ziegler_productivity_2022}.
Developers often accept AI-generated snippets with limited scrutiny, suggesting superficial quality control practices \citep{ziegler_productivity_2022, chen_evaluating_2021}.
Perceived productivity improvements do not always align with objective activity metrics, raising the risk of overestimating efficiency gains \citep{stray_developer_2025}.
Security vulnerabilities represent a further concern: coding assistants can be susceptible to prompt injection attacks, potentially leading to data contamination, unintended code execution, or data leakage \citep{liu_your_2025, cotroneo_vulnerabilities_2024, norton_how_2025}.

Long-term risks are more structural.
Continuous reliance on AI-generated suggestions may erode deep technical knowledge and increase dependency on AI systems \citep{ng_harnessing_2024,brynjolfsson_generative_2023,ding_codingcare_2025}.
Automatically generated code can introduce opaque dependencies and architectural inconsistencies, contributing to technical debt \citep{ebert_generative_2023,atif_celloai_2025, anderson_hidden_2025,moreschini_evolution_2026}.
Even syntactically correct code may contain subtle semantic or security-relevant weaknesses \citep{yetistiren_evaluating_2023,atif_celloai_2025}.
Organizational dependence on specific vendors poses additional strategic risks, including technology lock-in and compliance exposure \citep{ng_harnessing_2024}.

At the team level, some studies indicate reduced direct human interaction when AI tools mediate communication, code explanation or problem-solving \citep{ulfsnes_transforming_2024,cabrero-daniel_exploring_2024}, entailing risks of knowledge silos and diminished interpersonal trust.
To mitigate these effects, research recommends maintaining structured exchange formats such as code reviews and sprint retrospectives.
More broadly, governance demands increase as \ac{GenAI} becomes embedded in production workflows.
Issues of quality assurance, accountability, compliance, and traceability become more salient when AI-generated artifacts are incorporated into production systems \citep{geyer_case_2025,zhang_empowering_2024,ulfsnes_transforming_2024,diebold_backlogs_2025, bahi_integrating_2024}.
The literature emphasizes the need to formalize review obligations, clarify responsibility for AI-assisted outputs, and establish dedicated governance frameworks to prevent quality degradation and unmanaged risk exposure.

Several roadblocks constrain the realization of \ac{GenAI}'s full potential. A large-scale industry survey \citep{ahlawat_art_2024} shows that coding accounts for only 10--15 \% of total development time and less than 50 \% of developers use \ac{GenAI}. They conclude, that improvements in code generation alone have limited overall impact. To prevent automation risks from offsetting productivity gains, a human-in-the-loop approach (including quality review, domain judgment, and requirements oversight) is emphasized by industry reports \citep{deniz_unleash_2023, ahlawat_art_2024, struever_dev_2025}.

Overall, the literature indicates that \ac{GenAI} exerts measurable influence across all phases of the SDLC, with the strongest effects observed in design, implementation, and testing.
However, these gains are neither uniform nor automatically realized—short-term limitations in reliability and security, long-term risks such as skill erosion and technical debt as well as organizational roadblocks may constrain or offset productivity improvements.
Effective governance frameworks, structured human-in-the-loop validation, and deliberate organizational adaptation remain critical to translating GenAI's technical capabilities into lasting benefits.

\section{Survey Design}
\label{sec:design}

To complement the literature analysis with practitioner perspectives, we conduct a standardized online survey informed by the \ac{TAM} \citep{davis_perceived_1989, venkatesh_theoretical_2000}. TAM provides a widely used framework for examining how users perceive the usefulness and adoption of new technologies in organizational contexts.
The survey investigates how software practitioners use \ac{GenAI} tools, how they assess productivity effects across different \ac{SDLC} phases, how governance mechanisms for \ac{GenAI} adoption are implemented in practice, and which risks practitioners associate with the use of \ac{GenAI}.

Data were collected using a self-developed online questionnaire.
The instrument comprised five sections: (1) demographic and professional background variables (age, education, role, team size); (2) primary development methodology (agile, traditional, hybrid) and degree of adherence; (3) \ac{GenAI} tool usage behavior and frequency; (4) assessments of productivity, quality, and security implications across \ac{SDLC} phases; and (5) time-based efficiency estimations and open-ended questions on perceived opportunities and risks.
A pretest ($n = 3$) was conducted to refine wording and improve the distinction between process model categories.

The target population comprised professionals involved in software development and related disciplines, including software engineers, architects, team leads, DevOps engineers, product owners, QA engineers, as well as students and interns.
Data collection took place between 31 December 2025 and 19 January 2026 using a multi-stage snowball sampling approach \citep{doring_forschungsmethoden_2023} through professional and academic networks, including direct outreach on LinkedIn.
While the primary focus was the DACH region, participation from other regions was not restricted.

\section{Survey Results}
\label{sec:results}

In the following we present the findings of our quantitative online survey (n = 65).
Although the sample exhibits a comparatively young age structure, it demonstrates a strong professional profile.
The majority of respondents are industry practitioners, including 30 professional software developers, one team lead, and four product owners.
Notably, ten participants report more than ten years of professional experience, indicating that the sample includes senior expertise despite its youthful demographic composition.

\begin{figure}[ht]
  \centering
  \includegraphics[width=0.9\textwidth]{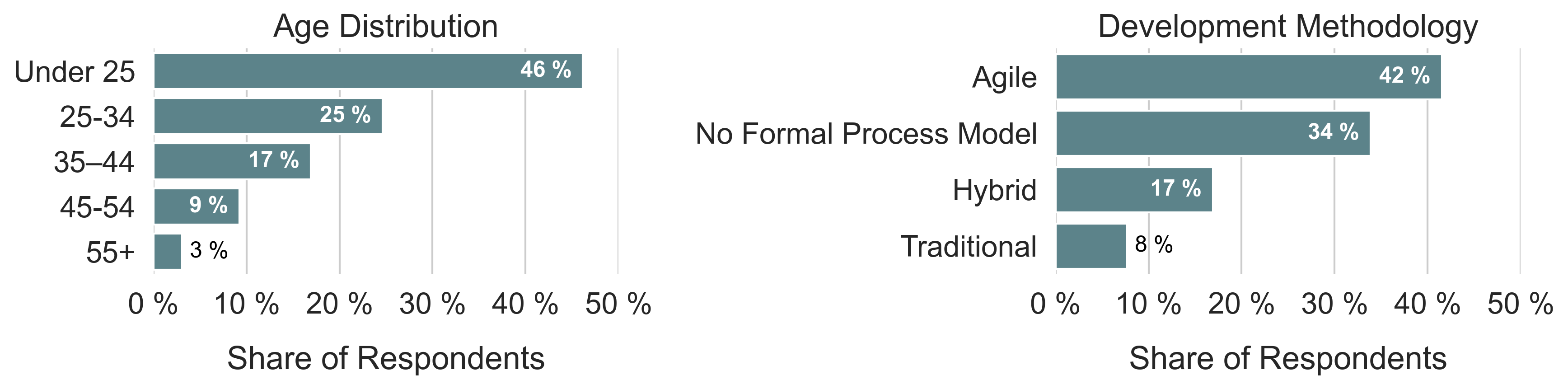}
  \caption{Development methodology and age distribution}
  \label{fig:dev_method_and_age}
\end{figure}

With regard to development methodologies, the agile approaches Scrum and Kanban constitute the largest share at about 42 \%. Approximately 34 \% report operating without a clearly defined process model, followed by 17 \% of respondents who use a hybrid approach with both agile and traditional (waterfall-based) methodologies.
The comparably smallest group with about 8 \% of respondents work in formalized or plan-driven (``Traditional'') environments.

\begin{figure}[ht]
  \centering
  \includegraphics[width=0.82\textwidth]{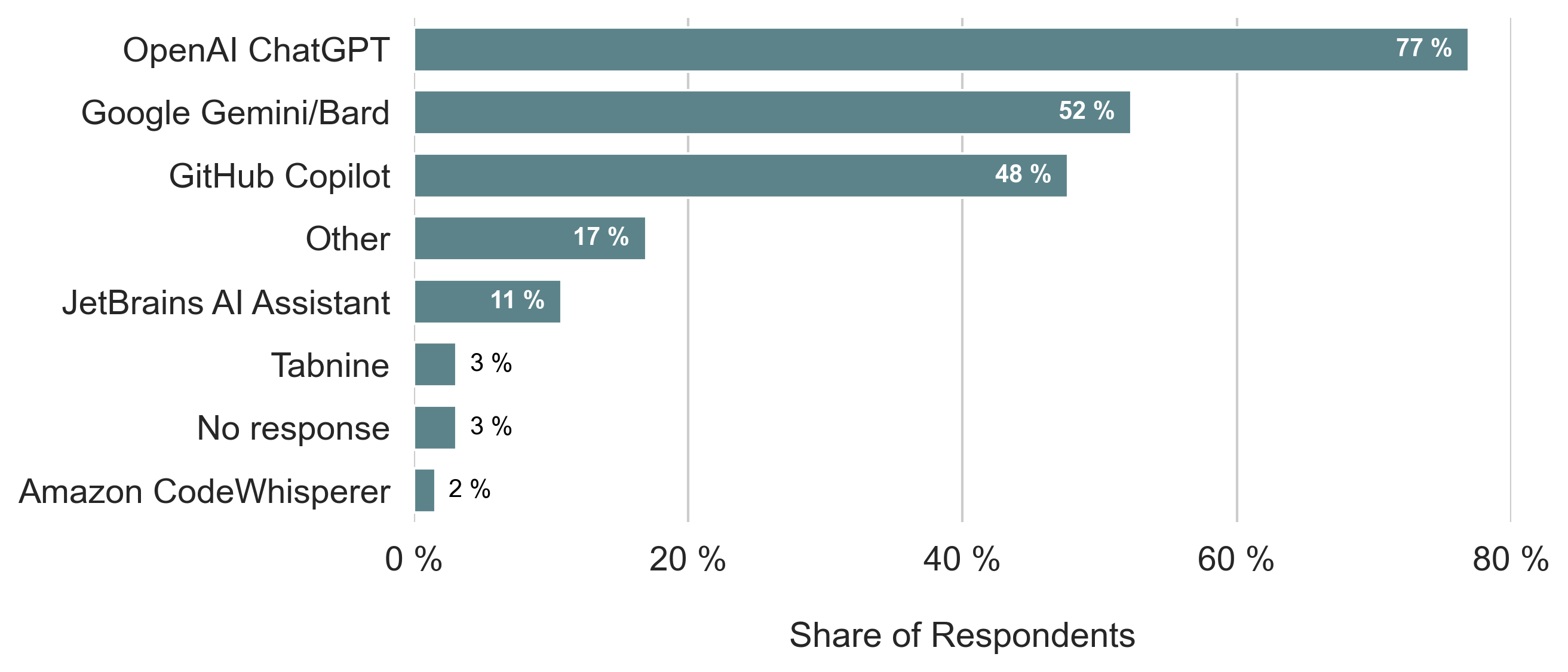}
  \caption{GenAI tool usage}
  \label{fig:tools}
\end{figure}

As Figure \ref{fig:tools} presents, the survey reveals widespread integration of \ac{GenAI} into daily development activities.
Approximately 79 \% of respondents report using AI tools at least once per day, indicating that \ac{GenAI} has become a routine component of professional practice.
In terms of tools, browser-based LLMs dominate. ChatGPT is used by 77 \% of participants, making it the most frequently adopted tool, followed by Google Gemini/Bard at 52 \%.
GitHub Copilot, which combines \ac{IDE}-integrated code completion with a conversational chat interface, is used by about 48 \% of respondents.
\ac{IDE}-integrated coding assistants such as JetBrains AI Assistant (11 \%), Tabnine (3 \%), and Amazon CodeWhisperer (2 \%) remain marginal by comparison. 17 \% of respondents reported using other tools while 3 \% did not respond to the tool-related question.
This distribution suggests that developers favor dialog-based interaction with general-purpose LLMs over specialized \ac{IDE}-integrated alternatives.

\begin{figure}[ht]
  \centering
  \includegraphics[width=0.88\textwidth]{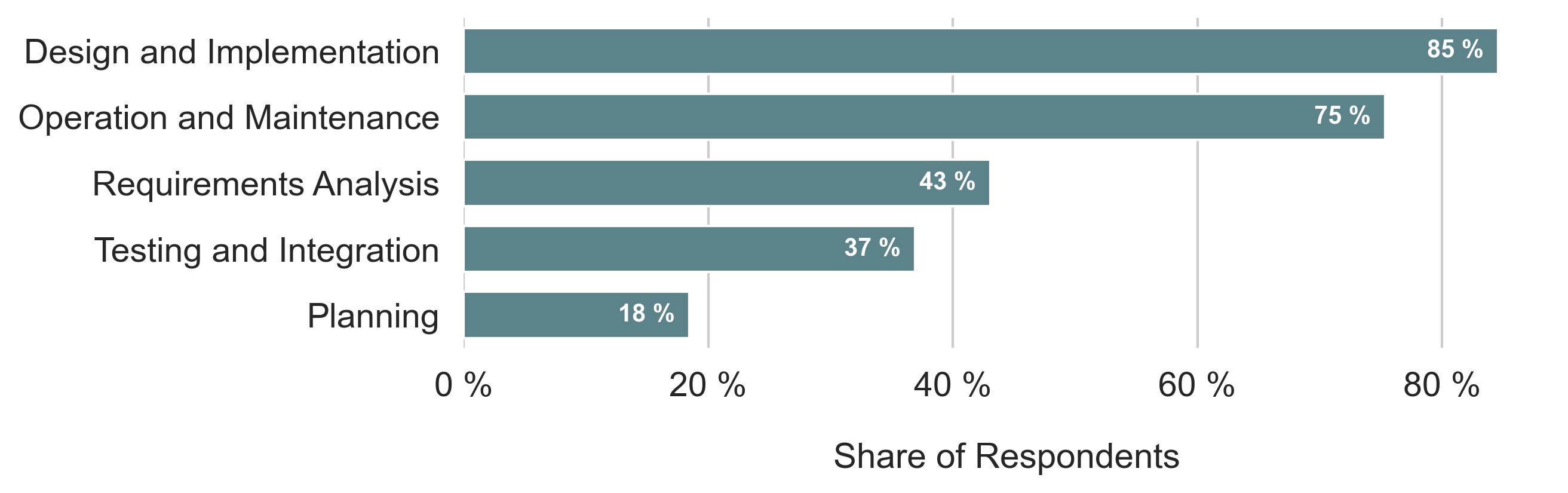}
  \caption{Impact of \ac{GenAI} by \ac{SDLC} phase}
  \label{fig:sdlc_phases}
\end{figure}

Figure \ref{fig:sdlc_phases} displays the distribution of responses regarding the \ac{SDLC} phases in which participants perceived the greatest utility of \ac{GenAI} (multiple choice).
85 \% of respondents identify “Design and Implementation” as the phase with the strongest perceived benefit, followed by “Operation and Maintenance” at 75 \%. Earlier lifecycle phases receive notably lower ratings, with ``Requirements Analysis'' at 43 \%, ``Testing and Integration'' at roughly 37 \%, and ``Planning'' at 18 \%.
Although the low value for ``Planning'' might be attributed to the high proportion of junior software developers in the sample, none of the Product Owners and only 5 \% of respondents with more than ten years of experience report a perceived impact in this phase, suggesting the opposite.

\begin{figure}[ht]
  \includegraphics[width=\textwidth]{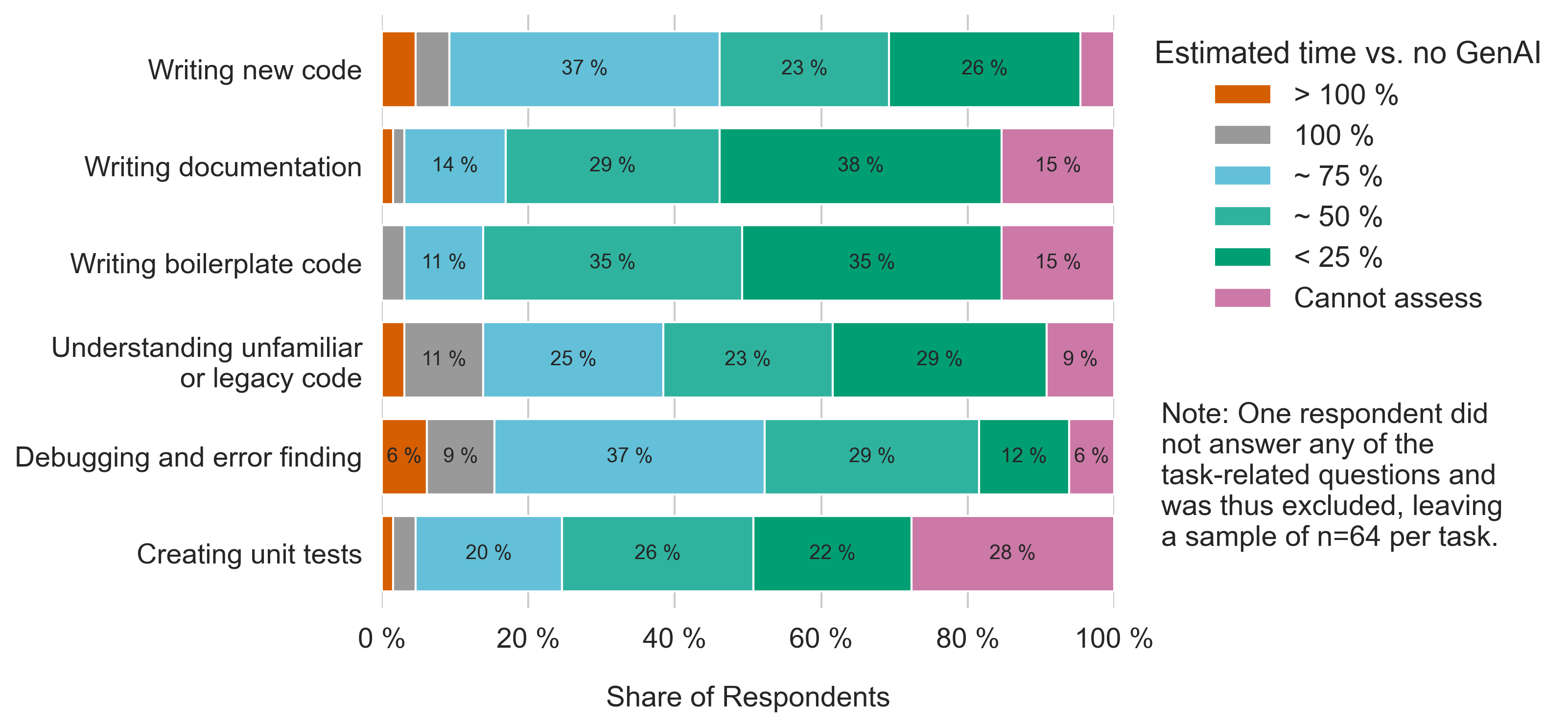}
  \caption{Estimated time saving with \ac{GenAI} by coding task}
  \centering(100 \% = no time saving, 0 \% = full automation)
  \label{fig:time_impact}
\end{figure}

To quantify these benefits at the task level, respondents were asked to estimate the time required for six common development activities when using \ac{GenAI} in relation to working without AI support (see Figure \ref{fig:time_impact}).
The results show that the strongest effects are reported for writing boilerplate code and documentation, where 72 \% and 69 \% of respondents, respectively, estimate at least halving the required time.
These two tasks fall under design and implementation as well as operation and maintenance; the two \ac{SDLC} phases also rated highest in Figure \ref{fig:sdlc_phases}.
Writing new code shows comparably more moderate positive effects, with the largest group (37 \%) reporting a notable reduction to approximately 75 \%.
Creating unit tests also shows substantial gains among the respondents, although the high “cannot assess” rate (28 \%) suggests that not all respondents regularly perform this activity.
Understanding unfamiliar or legacy code produces the most polarized responses, with 29 \% reporting extreme speedup to less than 25 \% but also the highest ``no change'' rate (11 \%) across all tasks.
Debugging yields the weakest perceived savings and the highest share of respondents reporting increased effort (6 \%), consistent with the cognitively complex nature of this task.
Together, those findings confirm the primary role of \ac{GenAI} as a coding assistant that accelerates repetitive and standardized development tasks, while also suggesting substantial perceived value in operation and maintenance activities such as debugging, log analysis, and legacy code comprehension.

\begin{figure}[ht]
  \centering
  \includegraphics[width=0.9\textwidth]{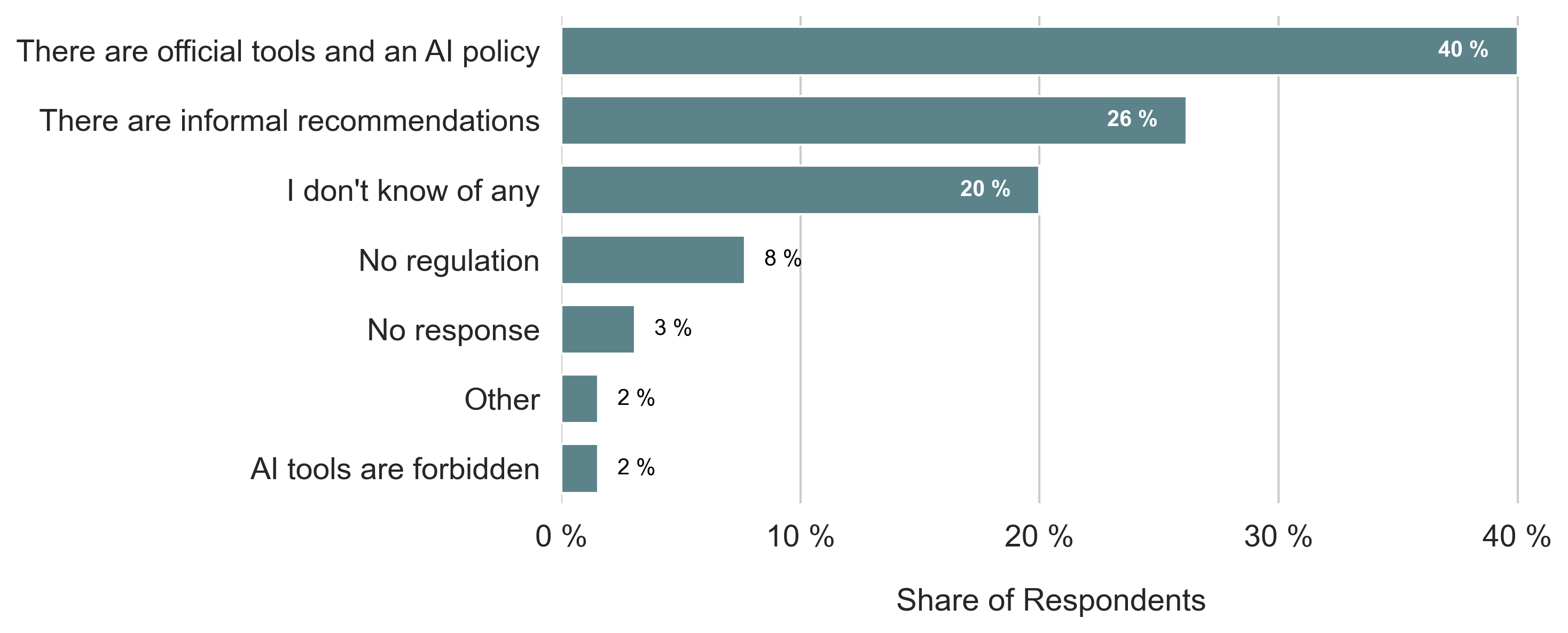}
  \caption{GenAI policies and guidelines in the respondent's companies}
  \label{fig:governance}
\end{figure}

As Figure \ref{fig:governance} shows, the survey results indicate a comparatively mature governance landscape for \ac{GenAI} usage across organizations.
The largest share of respondents (40 \%) report the existence of official tools combined with a formal AI policy.
An additional 26 \% indicate the presence of informal recommendations, suggesting that in roughly two thirds of organizations AI usage is at least partially embedded within recognized guidance structures.
However, governance is not uniformly institutionalized. Around 20 \% of participants state that they are unaware of any existing policies, and approximately 8 \% report that no regulation exists at all.
Smaller proportions indicate no response, other miscellaneous categories or that AI tools are explicitly forbidden in the company.
This distribution reveals a dual dynamic: While structured governance mechanisms are already widespread, a non-negligible segment of organizations still exhibits regulatory ambiguity or limited transparency regarding AI usage policies.

The qualitative responses largely reinforce the quantitative findings and provide deeper insight into perceived opportunities, risks, and anticipated role changes associated with \ac{GenAI} in software development.
Across roles and experience levels, the dominant theme is efficiency.
Participants repeatedly emphasize time savings through automation of repetitive and low-complexity tasks such as boilerplate code generation, documentation, test creation, debugging support, and code explanation.
Many respondents describe \ac{GenAI} as a “sparring partner” that accelerates learning, facilitates onboarding into new technologies, and supports rapid prototyping.
Experienced developers highlight its value in reviewing novel code, interfacing with unfamiliar systems, and exploring alternative solution paths. Several comments state that AI reduces ``toil'' and allows developers to focus more on architecture, system design, and complex problem solving.

At the same time, concerns are substantial and remarkably consistent.
The most frequently mentioned risks are uncritical adoption of AI-generated code, loss of deep technical understanding, and data protection and security vulnerabilities.
Many respondents warn against “blind copying” and emphasize that AI outputs often appear syntactically correct but may contain logical, architectural, or security flaws.
Senior practitioners stress that experience remains indispensable for evaluating AI-generated artifacts.
A recurring theme is the danger of cognitive offloading: if AI is used as a shortcut rather than as a support tool, individual learning curves and long-term problem-solving capabilities may deteriorate.
Data security---particularly the risk of exposing sensitive or company-specific information to external models---is also repeatedly cited as a major concern.

Regarding the future of the profession, most participants do not anticipate the disappearance of software development but rather a structural transformation of the role.
A common expectation is a shift from pure code writing toward architectural thinking, requirement formulation, quality assurance, and orchestration of AI systems. Several respondents predict a polarization effect: junior roles may decline in number, while senior and architect-level competencies become more valuable.
Others foresee new hybrid roles, such as AI supervisors.
While a minority expresses skepticism or uncertainty about long-term effects, the prevailing view is that \ac{GenAI} will become a standard tool, and that developers who fail to integrate it into their workflows may fall behind.

\section{Conclusion}
\label{sec:conclusion}

We examined the impact of \ac{GenAI} on the \ac{SDLC} by integrating a systematic literature review with empirical evidence from a cross-sectional survey of 65 software developers.
The findings consistently demonstrate that AI-assisted coding tools exert substantial influence across multiple \ac{SDLC} phases, with the most pronounced and empirically validated effects observed in design, implementation, testing, and documentation.
We determine that prior research reports measurable productivity gains, reductions in routine workload, and generally stable or improved code quality when human validation mechanisms remain in place.

The survey results largely corroborate these findings.
Respondents identify implementation, boiler plate code, and documentation as the domains with the highest perceived value, confirming the role of \ac{GenAI} as a productivity multiplier in syntactically structured and repetitive tasks.
In contrast, less structured and evaluative tasks such as planning, requirements analysis, debugging, and testing show considerably lower perceived productivity gains. Beyond phase-specific effects, the study highlights structural and organizational dimensions of AI integration.
\ac{GenAI} has become deeply embedded in daily workflows, yet security concerns remain moderately elevated, indicating that developers are aware of potential risks despite widespread adoption.
The governance analysis reveals that a majority of organizations have already established either formal AI policies or at least informal usage guidelines, yet, a meaningful minority report regulatory ambiguity or limited awareness of existing policies, pointing to communication and implementation gaps.

Importantly, the findings suggest that \ac{GenAI} does not eliminate the need for expertise but instead shifts value creation within the SDLC.
As routine coding activities become increasingly automated, higher-order competencies, such as requirements precision, architectural reasoning, critical validation, and governance oversight, gain importance.
Experience level moderates perception: early-career professionals report particularly strong perceived learning benefits, whereas experienced practitioners tend to frame AI primarily as an efficiency-enhancing tool.
This dynamic underscores the importance of structured training and human-in-the-loop mechanisms to prevent overreliance and ensure sustainable skill development.

Several limitations warrant consideration.
The survey sample is skewed toward younger, developer-centric respondents primarily from the DACH region.
Moreover, the reliance on self-reported productivity estimates introduces potential response biases, as perceived and objective productivity gains may diverge.

In summary, \ac{GenAI} is not merely an incremental productivity tool but a transformative force reshaping workflows, role profiles, and governance structures within software development.
However, the magnitude and durability of its benefits depend on deliberate organizational integration, transparent policy frameworks, and continued human accountability.
Promising future research avenues lie in examining longitudinal effects on skill development, architectural quality, and organizational performance.

\clearpage

% -----------------------------------------------------------------------------
% REFERENCES
% -----------------------------------------------------------------------------

\bibliographystyle{unsrtnat}
\bibliography{GenAI_Software_Dev}

% -----------------------------------------------------------------------------
% END OF DOCUMENT
% -----------------------------------------------------------------------------

\end{document}